\documentclass[11pt,onecolumn,oneside,letterpaper,peerreview]{IEEEtran}
\usepackage{cite}
\usepackage{graphicx}
\usepackage{amsmath}
\usepackage{times}
\usepackage{latexsym}
\usepackage{bm}
\usepackage{amssymb}
\usepackage[center]{caption2}
\usepackage{array}
\usepackage{setspace}
\usepackage{fancyhdr}
\usepackage{citesort}

\newtheorem{theorem}{Theorem}

\newcommand{\defeq}{\stackrel{\Delta}{=}}

\newcaptionstyle{mystyle2}{%
\captionlabel $.$ \, \doublespacing \captiontext \par}
\captionstyle{mystyle2}

\begin{document}
\title{Asymptotic SER and Outage Probability of MIMO MRC in Correlated Fading}
\author{Shi~Jin,~\IEEEmembership{Student Member,~IEEE,} Matthew R.
McKay, ~\IEEEmembership{Student Member,~IEEE,}\\
        Xiqi~Gao,~\IEEEmembership{Member,~IEEE,}
        and ~Iain B.~Collings,~\IEEEmembership{Senior~Member,~IEEE}
\thanks{Manuscript received April 18, 2006; revised June 4, 2006.
        The associate editor coordinating the review of this manuscript and approving it for publication was Prof. Zhengdao Wang.
        This work was supported by National Natural Science Foundation of China under Grants 60496311 and 60572072,
        the China High-Tech 863-FuTURE Project under Grant 2003AA123310, and the International Cooperation Project
        on Beyond 3G Mobile of China under Grant 2005DFA10360. }
\thanks{S. Jin and X. Gao are with the National Mobile Communication Research Laboratory,
Southeast University, Nanjing, 210096, P.R. China (email:
jinshi@seu.edu.cn; xqgao@seu.edu.cn).}
\thanks{M. R. McKay is with Telecommunications Lab., School of Elec.
and Info. Engineering, University of Sydney, Australia, and also
the ICT Centre, CSIRO, Australia (email: mckay@ee.usyd.edu.au).}
\thanks{I. B. Collings is with the Wireless Technologies Lab., ICT Centre,
CSIRO, Australia (e-mail: Iain.Collings@csiro.au).} }

\maketitle

\begin{abstract}
This letter derives the asymptotic symbol error rate (SER) and
outage probability of multiple-input multiple-output (MIMO) maximum
ratio combining (MRC) systems.  We consider Rayleigh fading channels
with both transmit and receive spatial correlation. Our results are
based on new asymptotic expressions which we derive for the p.d.f.\
and c.d.f.\ of the maximum eigenvalue of positive-definite quadratic
forms in complex Gaussian matrices. We prove that spatial
correlation does not affect the diversity order, but that it reduces
the array gain and hence increases the SER in the high SNR regime.
%
%
%
\end{abstract}

\vskip 2ex \vspace*{0.5cm}
\begin{tabular}{ll}
{EDICS Categpry\;:} {COM-MIMO MIMO communications and signal processing}\\
{Corresponding Author\;:} {Shi Jin}\\
{National Mobile Communication Research Laboratory,}\\
{Southeast University, Nanjing 210096, China}  \\
{E-mail: jinshi@seu.edu.cn} \\
\end{tabular}

\newpage
\section{Introduction}

Multiple-input multiple-output (MIMO) maximal ratio-combining (MRC)
systems have recently received much attention due to their ability
to mitigate the severe effects of fading through diversity
\cite{Lo99}. The performance of MIMO-MRC has been previously
investigated in various uncorrelated and
semi-correlated\footnote{channels with transmit or receive
correlation, but not both.} channel scenarios. The main performance
measures evaluated have been symbol error rate (SER) and outage
probability. Uncorrelated Rayleigh fading was considered in
\cite{Dighe03,Rao03,Kang04,Grant05,Chen05,Dai05}, and
semi-correlated Rayleigh fading was considered in
\cite{kang03_icc,Zanella05,McKay06}.

In this paper we present results for double-correlated channels.
In particular, we derive new asymptotic expressions for the SER
and outage probability for arbitrary numbers of antennas, which
are simple functions of the system and channel parameters.  These
expressions are particularly useful in comparison to the only
other known double-correlated MIMO-MRC results, presented in
\cite{McKay05}, which in any case had SER expressions limited to
$2$-antenna systems\footnote{systems with two antennas at either
the transmitter or the receiver.} . Moreover, the results we
present here collapse trivially to semi-correlated scenarios, and
provide simpler expressions than the previous results in
\cite{kang03_icc,Zanella05,McKay06}.

Our results hinge on new first-order expansions which we derive
for the probability density function (p.d.f.) and cumulative
distribution function (c.d.f.) of the maximum eigenvalue of
positive-definite quadratic forms in complex Gaussian matrices.
From these, we obtain a high SNR expression for the SER, and
determine the diversity order and array gain. In so doing, we
prove that MIMO-MRC achieves the maximum spatial diversity order,
and we quantify the increase in SER in the high SNR regime due to
spatial correlation. We also analyze the outage probability at
outage levels of practical interest, and again quantify the effect
of spatial correlation.

\section{MIMO-MRC System Model}\label{sec:model}

Consider an $N_r  \times N_t$ MIMO system, where the $N_r \times
1$ received signal vector is
\begin{equation}\label{eq:sigmodel}
{\mathbf{r}} = \sqrt {\bar \gamma } {\mathbf{Hw}}x + {\mathbf{n}},
\end{equation}
where $x$ is the transmitted symbol with $E\left[ {\left| x
\right|^2 } \right] = 1$, $\mathbf{w}$ is the beamforming (BF)
vector with $E\left[ {\left\| {\mathbf{w}} \right\|^2 } \right] =
1$, ${\mathbf{n}} \in \mathcal{C}^{N_r  \times 1}$ is additive
noise vector $ \sim \mathcal{CN}_{N_r ,1} \left(
{{\mathbf{0}}_{N_r \times 1} ,{\mathbf{I}}_{N_r } } \right)$, and
$\bar \gamma $ is the transmit SNR. Also, $\mathbf{H}$ is the $N_r
\times N_t$ channel matrix, assumed to be flat
spatially-correlated Rayleigh fading.

We assume that $\mathbf{H}$ can be decomposed according to the
popular Kronecker correlation structure (as in
\cite{Shiu00,McKay06,kang03_icc}) as follows
\begin{equation}\label{eq:channel}
{\mathbf{H}} = {\mathbf{R}}^{\frac{1} {2}} {\mathbf{H}}_w
{\mathbf{S}}^{\frac{1} {2}},
\end{equation}
where $\mathbf{R} > 0$ and $\mathbf{S} > 0$ are the receive and
transmit spatial correlation matrices respectively, with unit
diagonal entries, and ${\mathbf{H}}_w  \sim \mathcal{CN}_{N_r ,N_t }
\left( {{\mathbf{0}}_{N_r \times N_t } ,{\mathbf{I}}_{N_r }  \otimes
{\mathbf{I}}_{N_t } } \right)$.

The receiver employs the principle of MRC to give
\begin{align}\label{eq:recMRC}
{z} = {\mathbf{w}}^\dag  {\mathbf{H}}^\dag  {\mathbf{r}} = \sqrt
{\bar \gamma } {\mathbf{w}}^\dag  {\mathbf{H}}^\dag {\mathbf{Hw}}x +
{\mathbf{w}}^\dag  {\mathbf{H}}^\dag {\mathbf{n}}.
\end{align}
Therefore, the SNR at the output of the combiner is easily derived
as
\begin{align}\label{eq:SNRcom}
\gamma  = \bar \gamma {\mathbf{w}}^\dag  {\mathbf{H}}^\dag
{\mathbf{Hw}}.
\end{align}
The BF vector $\mathbf{w}$ is chosen to maximize this instantaneous
output SNR, thereby minimizing the error probability. It is well
known that the optimum BF vector ${\mathbf{w}}_{{\rm opt}}$ is the
eigenvector corresponding to the maximum eigenvalue $\lambda _{ {\rm
max}}$ of ${\mathbf{H}}^\dag {\mathbf{H}}$. In this case, the output
SNR (\ref{eq:SNRcom}) becomes
\begin{align}\label{eq:optvec}
\gamma  = \bar \gamma {\mathbf{w}}_{{\rm opt}}^\dag
{\mathbf{H}}^\dag {\mathbf{Hw}}_{{\rm opt}}  = \bar \gamma \lambda
_{{\rm max}}.
\end{align}

Clearly the performance of MIMO-MRC depends directly on the
statistical properties of $\lambda _{{\rm max}}$. For
double-correlated Rayleigh fading channels, $\lambda _{{\rm max}}$
is statistically equivalent to the maximum eigenvalue of a
positive-definite quadratic form in complex Gaussian random matrices
${\mathbf{Y}} \sim \mathcal{Q}_{n,m} \left( {{\mathbf{I}}_n
,{\mathbf{\Omega }},{\mathbf{\Sigma }}} \right)$ (see \cite{Gupta00}
and \cite{Shin03} for more details), where $m = \max \left( {N_r
,N_t } \right)$, $n = \min \left( {N_r ,N_t } \right)$, and
${\mathbf{\Omega }} \in \mathcal{C}^{n \times n}$ and
${\mathbf{\Sigma }} \in \mathcal{C}^{m \times m}$ are the Hermitian
positive-definite matrices
\begin{align}\label{eq:sigphi}
{\mathbf{\Omega }} \triangleq \left\{ {\begin{array}{*{20}c}
   {\mathbf{R}} & {N_r  \leqslant N_t }  \\
   {\mathbf{S}} & {N_r  > N_t }  \\
 \end{array} } \right. \hspace*{2cm} {\mathbf{\Sigma }} \triangleq \left\{ {\begin{array}{*{20}c}
   {\mathbf{S}} & {N_r  \leqslant N_t }  \\
   {\mathbf{R}} & {N_r  > N_t }  \\
 \end{array} } \right.
\end{align}
with eigenvalues $\omega _1  <  \ldots < \omega _n$ and $\sigma _1 <
\ldots  < \sigma _m$ respectively.


\section{Maximum Eigenvalue Distribution of a Quadratic Form in Complex Gaussian Matrices}\label{sec:quadratic}

We now present a theorem which gives new first-order expansions for
the p.d.f.\ and c.d.f.\ of the maximum eigenvalue of a quadratic
form in complex Gaussian matrices.  This will allow us to derive the
asymptotic distribution of the SNR in (\ref{eq:optvec}).

\begin{theorem}\label{theorem1}
The following two expressions are first order expansions of the
c.d.f. and p.d.f. respectively, of the maximum eigenvalue $\lambda
_{\rm max}$ of ${\mathbf{Y}} \sim \mathcal{Q}_{n,m} \left(
{{\mathbf{I}}_n ,{\mathbf{\Omega }},{\mathbf{\Sigma }}} \right)$:
\begin{align}\label{eq:cdffinal}
F_{\lambda _{\rm max} } \left( x \right) = \alpha x^{mn}  + o\left(
{x^{mn} } \right)
\end{align}
and
\begin{align}\label{eq:pdffinal}
f_{\lambda _{\rm max} } \left( x \right) = m n \alpha x^{mn - 1}
+ o\left( {x^{mn - 1} } \right),
\end{align}
where
\begin{align}\label{eq:alpha}
\alpha {\text{ = }} \frac{{\Gamma _n \left( n \right)}} {{\det
\left( {\mathbf{\Omega }} \right)^m \det \left( {\mathbf{\Sigma }}
\right)^n  \Gamma_n(m+n)}}
\end{align}
and $\Gamma _\cdot \left( \cdot \right)$ is the normalized complex
multivariate gamma function defined as
\begin{align}
\Gamma _n \left( m \right) = \prod\limits_{i = 1}^n {\Gamma \left(
{m - i + 1} \right)}.
\end{align}
\end{theorem}
\begin{proof}
The exact c.d.f. of the maximum eigenvalue of a positive-definite
quadratic form in complex Gaussian matrices is given by
\cite[Theorem 1]{McKay05}
\begin{align}\label{eq:cdf}
F_{\lambda _{\rm max} } \left( x \right) = \frac{{\left( { - 1}
\right)^n \Gamma _n \left( n \right)\det \left( {\mathbf{\Omega }}
\right)^{n - 1} \det \left( {\mathbf{\Sigma }} \right)^{m - 1} \det
\left( {{\mathbf{\Psi }}\left( x \right)} \right)}} {{\Delta _n
\left( {\mathbf{\Omega }} \right)\Delta _m \left( {\mathbf{\Sigma }}
\right)\left( { - x} \right)^{n\left( {n - 1} \right)/2} }},
\end{align}
where $\Delta _m \left(  \cdot  \right)$ is a Vandermonde
determinant in the eigenvalues of the \textit{m}-dimensional
matrix argument, given by
\begin{align} \label{eq:VandDef}
\Delta _m \left( {\mathbf{\Sigma }} \right) = \left| \sigma_i^{j-1}
\right| = \prod\limits_{i < j}^m {\left( {\sigma _j  - \sigma _i }
\right)}.
\end{align}\label{eq:phix}
Also, ${{\mathbf{\Psi }}\left( x \right)}$ is an $m \times m$
matrix with $\left( {i,j} \right)^{th}$ element
\begin{align}\label{eq:phi1}
\left\{ {{\mathbf{\Psi }}\left( x \right)} \right\}_{i,j}  =
\left\{ {\begin{array}{*{20}c}
   {\left( {\frac{1}
{{\sigma _j }}} \right)^{m - i} } & {i \leqslant \tau }  \\
   {e^{ - \frac{x}
{{\omega _{i - \tau } \sigma _j }}} P\left( {m; - \frac{x}
{{\omega _{i - \tau } \sigma _j }}} \right)} & {i > \tau }  \\
 \end{array} } \right.
\end{align}
where $\tau  = m - n$, and
\begin{align}\label{eq:ply}
P\left( {l;y} \right) = 1 - e^{ - y} \sum\limits_{k = 0}^{l - 1}
{y^k /k!}
\end{align}
is the regularized lower incomplete gamma function.

We seek a first-order expansion of (\ref{eq:cdf}) which will be
immediate from a first order expansion of $\det \left(
{{\mathbf{\Psi }}\left( x \right)} \right)$. Consider the Taylor
expansion of $\det \left( {{\mathbf{\Psi }}\left( x \right)}
\right)$ around the origin
\begin{align}\label{eq:fxtaylor}
\det \left( {{\mathbf{\Psi }}\left( x \right)} \right) \; = \;
F\left( x \right) \; = \; \sum\limits_{q = 0}^Q {\frac{{F^{\left(
q\right)} \left( 0 \right)}} {{q!}}} x^q  + o\left( {x^Q } \right).
\end{align}
For a first order expansion we need to find the first non-zero
coefficient in the sum. Using a well-known result for the $q$th
derivative of a determinant, we have
\begin{align}\label{eq:fxdiff}
F^{\left( q \right)} (0) = \sum\limits_{\left\{ {q_1 , \ldots ,q_m
} \right\}}^{} {\frac{{q!}} {{q_1 ! \cdots q_m !}}} \det \left(
\frac{{d^{q_i } \left\{ {{\mathbf{\Psi }}\left( x \right)}
\right\}_{i,j} }} {{dx^{q_i } }} \right) \biggr|_{x=0},
\end{align}
where $q_1  + \cdots  + q_m = q$. Using (\ref{eq:phi1}) and
(\ref{eq:ply}), we evaluate the derivatives in (\ref{eq:fxdiff})
as follows
\begin{equation}\label{eq:psi1}
\frac{{d^{q_i } \left\{ {{\mathbf{\Psi }}\left( x \right)}
\right\}_{i,j} }} {{dx^{q_i } }}  \biggr|_{x=0} = \left\{
{\begin{array}{cl}
   {\left( {\frac{1}
{{\sigma _j^{m - i} }}} \right) } & {i \leqslant \tau } \; \text{and } \; q_i = 0  \\
0 & {i \leqslant \tau } \; \text{and } \;  q_i > 0  \\
0 & {i > \tau } \; \text{and } \; 0 \leq q_i < m    \\
   {\left( { - \frac{1}
{{\omega _{i - \tau } \sigma _j }}} \right)^{ {q_i }} } & {i > \tau } \; \text{and } \; q_i \geq m \\
 \end{array} } \right.
\end{equation}
From (\ref{eq:psi1}), in order for the determinants in
(\ref{eq:fxdiff}) to be nonzero, we require that for rows $i
\leqslant \tau$ we must have $q_i = 0$ . We also require that for
rows $i > \tau$, we must have $q_i \geq m$, and these $q_i$'s must
all be different. Hence the \emph{smallest} $q$ for which these
conditions are satisfied is given by
\begin{align} \label{eq:qmin}
\tilde{q}&= m + (m+1) + \ldots + (m + (m-\tau-1)) \nonumber \\
&= mn + n (n-1)/2.
\end{align}
For this smallest $q$ value, we can now write (\ref{eq:fxdiff}) as
follows
\begin{align}\label{eq:fxdiff_min}
F^{\left( \tilde{q} \right)} (0) = \sum\limits_{\left\{
\underline{\alpha} \right\}}^{} {\frac{{\tilde{q}!}} {\Gamma _n
\left( {m + n} \right) }} \det \left( \frac{{d^{\tilde q_i }
\left\{ {{\mathbf{\Psi }}\left( x \right)} \right\}_{i,j} }}
{{dx^{\tilde q_i } }}
 \right) \biggr|_{x=0},
\end{align}
where the sum is over all permutations $\underline{\alpha} = \left\{
{\alpha_{1} , \ldots , \alpha_n } \right\}$ of the numbers $\left\{
{0, \ldots ,n - 1} \right\}$, and
\begin{align} \label{eq:qi}
\tilde{q}_i  = \left\{ {\begin{array}{*{20}c}
   0 & {i \leqslant \tau }  \\
   {m + \alpha_{i - \tau } } & {i > \tau }  \\
 \end{array} } \right.
\end{align}
Using (\ref{eq:psi1}) in (\ref{eq:fxdiff_min}) we have
\begin{align}\label{eq:fxdiff_min1}
F^{\left( \tilde{q} \right)} (0) =  {\frac{{\tilde{q}!} \;
(-1)^{\tilde{q}}} {\Gamma _n \left( {m + n} \right) }}
\sum\limits_{\left\{ \underline{\alpha} \right\}}^{} \det \left(
\mathbf{\Xi}_{\underline{\alpha}} \right)
\end{align}
%
where
\begin{equation}\label{eq:Xi}
\left\{ {\mathbf{\Xi }}_{\underline{\alpha}} \right\}_{i,j}  =
\left\{ {\begin{array}{*{20}c}
   {\left( {\frac{1}
{{\sigma _j }}} \right)^{m - i} } & {i \leqslant \tau }  \\
   {\left( {\frac{1}
{{\omega _{i - \tau } \sigma _j }}} \right)^{m + \alpha_{i - \tau } } } & {i > \tau }  \\
 \end{array} } \right.
\end{equation}
We now focus on simplifying the determinant sum in
(\ref{eq:fxdiff_min1}).  By removing the factors $\left( {1/\omega
_{i - \tau } } \right)^{m + \alpha_{i-\tau} }$ from the
determinants, and performing some row swaps, it can be shown that
\begin{align}
\sum\limits_{\left\{ \underline{\alpha} \right\}}^{} \det \left(
\mathbf{\Xi}_{\underline{\alpha}} \right) &= \sum\limits_{\left\{
\underline{\alpha} \right\}}^{} \left( \prod_{i=\tau+1}^m \left(
\frac{1}{\omega_{i-\tau}} \right)^{m + \alpha_{i-\tau}} (-1)^{{\rm
per}(\underline{\alpha})} (-1)^{\tau(\tau-1)/2} \det \left( \left(
\frac{1}{\sigma_i} \right)^{n+j-1} \right) \right) \nonumber
\\
&= \frac{ (-1)^{\tau(\tau-1)/2} \det \left( \left(
\frac{1}{\sigma_i} \right)^{j-1} \right)}{\det \left(
\mathbf{\Omega} \right)^m \det \left( \mathbf{\Sigma} \right)^n }
\sum\limits_{\left\{ \underline{\alpha} \right\}}^{} (-1)^{{\rm
per}(\underline{\alpha})} \prod_{i=1}^n
\left(\frac{1}{\omega_i}\right)^{\alpha_i} \nonumber \\
&= \frac{ (-1)^{\tau(\tau-1)/2} \det \left( \left(
\frac{1}{\sigma_i} \right)^{j-1} \right) \det \left( \left(
\frac{1}{\omega_i} \right)^{j-1} \right)}{\det \left(
\mathbf{\Omega} \right)^m \det \left( \mathbf{\Sigma} \right)^n }
\label{eq:sumDets}
\end{align}
where the last line followed from the definition of the determinant.
Next, using (\ref{eq:VandDef}) and the Vandermonde determinant
identity \cite[Eq.\ (56)]{McKay05}
\begin{align}
\prod_{i<j}^m \left( \frac{1}{\sigma_j} - \frac{1}{\sigma_i}
\right) = \frac{{\prod\nolimits_{i < j}^m {\left( {\sigma _i  -
\sigma _j } \right)} }} {{\prod\nolimits_{i = 1}^m {\sigma _i^{m -
1} } }},
\end{align}
we can write (\ref{eq:sumDets}) as follows
\begin{align}
\sum\limits_{\left\{ \underline{\alpha} \right\}}^{} \det \left(
\mathbf{\Xi}_{\underline{\alpha}} \right) &= (-1)^{\tau(\tau-1)/2}
(-1)^{m(m-1)/2} (-1)^{n(n-1)/2} \frac{ \Delta_m \left(
\mathbf{\Sigma} \right) \Delta_n \left( \mathbf{\Omega} \right) }{
\det \left( \mathbf{\Sigma} \right)^{m+n-1} \det \left(
\mathbf{\Omega} \right)^{m+n-1} } \nonumber \\
&= (-1)^{n(m+1)} \frac{ \Delta_m \left( \mathbf{\Sigma} \right)
\Delta_n \left( \mathbf{\Omega} \right) }{ \det \left(
\mathbf{\Sigma} \right)^{m+n-1} \det \left( \mathbf{\Omega}
\right)^{m+n-1} } \; \; . \label{eq:alphaSum_Final}
\end{align}
Substituting (\ref{eq:alphaSum_Final}) into (\ref{eq:fxdiff_min1})
and using (\ref{eq:qmin}), we derive the desired first-order
expansion of $\det \left( {{\mathbf{\Psi }}\left( x \right)}
\right)$ in (\ref{eq:fxtaylor}), given by
\begin{align} \label{eq:detFirstOrder}
\det \left( {{\mathbf{\Psi }}\left( x \right)} \right) = \frac{
(-1)^n (-1)^{n(n-1)/2} \Delta_m \left( \mathbf{\Sigma} \right)
\Delta_n \left( \mathbf{\Omega} \right) }{\Gamma_{n}(m+n) \det
\left( \mathbf{\Sigma} \right)^{m+n-1} \det \left( \mathbf{\Omega}
\right)^{m+n-1} } x^{mn + n(n-1)/2} + o (x^{mn + n(n-1)/2}) \; .
\end{align}
The c.d.f.\ result (\ref{eq:cdffinal}) now follows by substituting
(\ref{eq:detFirstOrder}) into (\ref{eq:cdf}) and simplifying.  The
p.d.f.\ result (\ref{eq:pdffinal}) then follows trivially by taking
the derivative of (\ref{eq:cdffinal}) w.r.t.\ $x$.
\end{proof}
For the special case of uncorrelated fading, (\ref{eq:cdffinal})
reduces to
\begin{equation}\label{eq:cdfuncorre}
F_{\lambda _{max} } \left( x \right) = \frac{{\Gamma _n \left( n
\right)}} { {\Gamma_n(m+n) } }x^{mn}  + o\left( {x^{mn} } \right).
\end{equation}
which agrees with a result derived previously in \cite{Dai05}.

\section{Asymptotic SER and Outage Analysis of MIMO-MRC}\label{sec:analysis}

For many general modulation formats, the average SER of MIMO-MRC
can be expressed as
\begin{align}\label{eq:serexp}
P_s  = E_\gamma  \left[ {aQ\left( {\sqrt {2b\gamma } } \right)}
\right],
\end{align}
where $Q\left(  \cdot  \right)$ is the Gaussian Q-function, and
$a$ and $b$ are modulation-specific constants \cite{Proakis01}. In
\cite{Chen04}, an alternative expression of (\ref{eq:serexp}) was
provided as follows
\begin{align}\label{eq:serexp1}
P_s  = \frac{{a\sqrt b }} {{2\sqrt \pi  }}\int_0^\infty
{\frac{{e^{ - bu} }} {{\sqrt u }}} F_{\lambda _{\max } } \left( u
\right)du.
\end{align}

We now analyze the SER performance in the high SNR regime in order
to derive the diversity order and array gain of the system. Armed
with \textit{Theorem \ref{theorem1}}, we can directly invoke a
general parameterized single-input single-output (SISO) SER result
from \cite{Wang03}, and perform some basic algebraic manipulations,
to obtain a high SNR SER expression given by
\begin{align}\label{eq:serhigh}
{\text{SER}}^\infty   = \left( {G_a  \cdot \bar \gamma } \right)^{
- G_d }  + o\left( {\bar \gamma ^{ - G_d } } \right),
\end{align}
where the diversity order is
\begin{align}\label{eq:diversity}
G_d  = mn
\end{align}
and the array gain is
\begin{align}\label{eq:array}
G_a  =  \det \left( \mathbf{\Omega} \right)^{1/n} \det \left(
\mathbf{\Sigma} \right)^{1/m} 2 b \left(  \frac{ a \Gamma_n(n)}{2
\Gamma_n(m+n)} (2 m n - 1)!!
  \right)^{ - 1/mn}
\end{align}
with
\begin{align}
\left( 2mn - 1 \right)!! \defeq 1 \times 3 \times \ldots \times
(2mn-1) \; .
\end{align}
We clearly see that MIMO-MRC achieves the full spatial diversity
order of $m n$, regardless of the spatial correlation.
Moreover, using Hadamard's
inequality and the fact that the diagonal elements of
$\mathbf{\Omega}$ and $\mathbf{\Sigma}$ are unity, it is easily
found that
\begin{align}\label{ineq:corr}
0 \leqslant \det \left( {\mathbf{\Omega }} \right) \leqslant 1
\hspace*{0.3cm} {\text{  and  }} \hspace*{0.3cm} 0 \leqslant \det
\left( {\mathbf{\Sigma }} \right) \leqslant 1
\end{align}
with equality in the upper limit only when the correlation
matrices are identity matrices. Hence, from (\ref{eq:array}) we
see that the effect of the correlation is to reduce the array gain
(with respect to uncorrelated fading) by a factor of $\det \left(
\mathbf{\Omega} \right)^{1/n} \det \left( \mathbf{\Sigma}
\right)^{1/m}$, thereby increasing the SER in the high SNR regime.
Note that for the special case $n = 2$, this result can be shown
to reduce to an expression reported previously in \cite{McKay05}.

We now consider the outage probability of MIMO-MRC systems in
double-correlated Rayleigh channels. The outage probability is an
important quality of service measure, defined as the probability
that $\gamma$ drops below an acceptable SNR threshold $\gamma_{{\rm
th}}$. It is obtained using (\ref{eq:optvec}) as follows
\begin{align}\label{eq:outage}
F_\gamma  \left( {\gamma _{\rm th} } \right) = \Pr \left( {\gamma
\leqslant \gamma _{\rm th} } \right) = F_{\lambda _{max}} \left(
{\frac{{\gamma _{\rm th} }} {{\bar \gamma }}} \right).
\end{align}
In practice, we are usually interested in small outage
probabilities (i.e. 0.01, 0.001, ...), which correspond to small
values of $\gamma_{\rm th}$. To gain further intuition at these
small outage probabilities, we use (\ref{eq:cdffinal}) in
\textit{Theorem {\ref{theorem1}}} to write the outage probability
in (\ref{eq:outage}) as follows
\begin{align}\label{eq:outagehigh}
\tilde F_\gamma  \left( {\gamma _{\rm th} } \right) = \frac{{\Gamma
_n \left( n \right)}} {{\det \left( {\mathbf{\Omega }} \right)^m
\det \left( {\mathbf{\Sigma }} \right)^n  \Gamma_n(m+n)}} \left(
{\frac{{\gamma _{\rm th} }} {{\bar \gamma }}} \right)^{mn}  +
o\left( {\left( {\gamma _{\rm th} } \right)^{mn} } \right).
\end{align}
According to (\ref{ineq:corr}), this result shows explicitly that in
the low outage regime the outage performance degrades due to the
presence of spatial correlation.  Moreover, the increase in outage
probability (with respect to uncorrelated fading) is quantified by
the factor $\det \left( {\mathbf{\Omega }} \right)^{-m} \det \left(
{\mathbf{\Sigma }} \right)^{-n}$.

\section{Numerical Results}\label{sec:simulation}

For our numerical results we construct the correlation matrices
using the exponential correlation model. The $\left( {i,j}
\right)$th entries of ${\mathbf{\Omega }}$ and ${\mathbf{\Sigma
}}$ are given by $\left\{ {{\mathbf{\Omega }}} \right\}_{i,j} =
\rho _1^{\left| {i - j} \right|}$ and $\left\{ {{\mathbf{\Sigma }}
} \right\}_{i,j} =  \rho _2^{\left| {i - j} \right|}$ with $\rho
_1 ,\rho _2 \in \left[ {0,1} \right)$, respectively. Note however,
that all of the analytical results presented in this paper apply
equally to any other correlation model that conforms to the
general structure in (\ref{eq:channel}).

\begin{figure}
\centering
\includegraphics[scale=0.7]{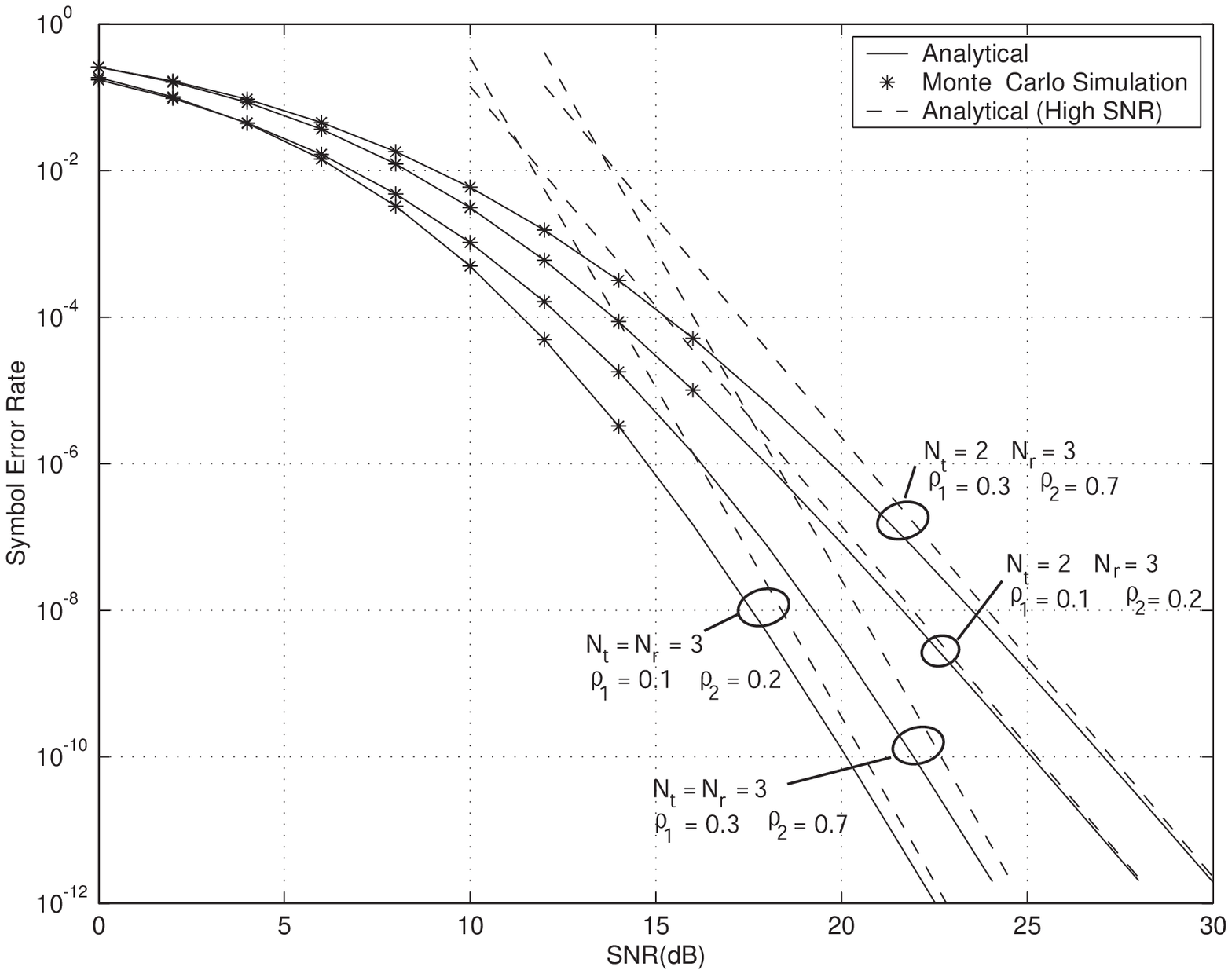}
\captionstyle{mystyle2}\caption{Symbol Error Rate of MIMO-MRC in
various double-correlated Rayleigh channels with 8PSK modulation.}
\label{fig:fig1}
\end{figure}

Fig.\ \ref{fig:fig1} shows the SER of MIMO-MRC with 8PSK ($a = 2$,
$b = 0.146$) modulation, for various antenna configurations. The
'Analytical' curves are generated via numerical integration of
(\ref{eq:serexp1}) using the analytical c.d.f.\ (\ref{eq:cdf}).
The 'Analytical (High SNR)' curves are based on
(\ref{eq:serhigh}). Clearly the diversity orders and array gains
predicted by the high SNR analytical results are accurate. Also,
as expected from the analysis in Section \ref{sec:analysis}, we
see that the SER increases monotonically with the level of
correlation for both antenna configurations.

\begin{figure}
\centering
\includegraphics[scale=0.7]{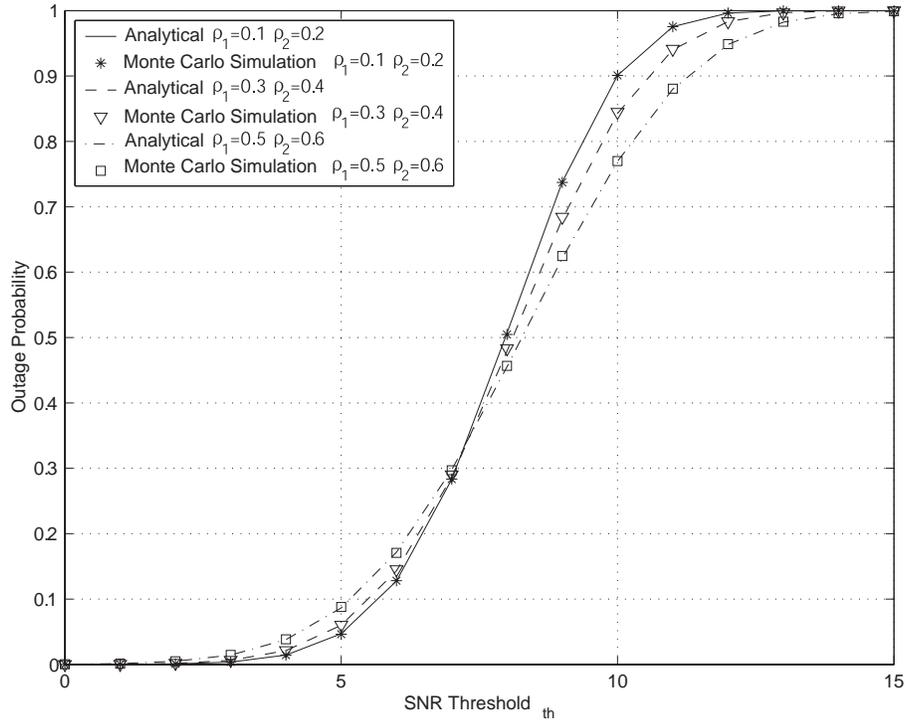}
\captionstyle{mystyle2}\caption{Outage probability of $3 \times 3$
MIMO-MRC in various double-correlated Rayleigh channels, and for
$\bar \gamma = 0{\rm dB}$.} \label{fig:fig2}
\end{figure}

Fig.\ \ref{fig:fig2} shows analytical and Monte-Carlo simulation
outage probability curves for a MIMO-MRC system, comparing different
correlation scenarios. The analytical results are based on
(\ref{eq:cdf}). 
As expected from our asymptotic analysis in Section
\ref{sec:analysis}, we see that the correlation increases the outage
probability for all outage levels of practical interest (i.e.\ in
this case, for outage levels $< 30\%$). It is also interesting to
observe that the opposite occurs for high outage levels.

\section{Conclusions}\label{sec:conclusion}

We have examined the asymptotic performance of MIMO-MRC systems in
double-correlated Rayleigh channels. Our results are based on new
closed-form asymptotic expressions which we have derived for the
marginal maximum eigenvalue distribution of positive-definite
quadratic forms in complex Gaussian matrices. The new results
prove that the presence of spatial correlation yields a net
increase in SER in the high SNR regime, and also degrades the
outage performance for outage levels of practical interest.


\newpage

\end{document}